\documentclass[a4paper,fleqn,usenatbib,useAMS]{mnras}

\usepackage{graphicx}   
\usepackage{amsmath}    
\usepackage{amssymb}    
\usepackage{multicol}   
\usepackage{bm}         
\usepackage{pdflscape}  
\def\la{\;
\raise0.3ex\hbox{$<$\kern-0.75em\raise-1.1ex\hbox{$\sim$}}\; }
\def\ga{\;
\raise0.3ex\hbox{$>$\kern-0.75em\raise-1.1ex\hbox{$\sim$}}\; }

\newcommand{\dmm}{$\Delta\mu/\mu$}

\newcommand{\kms}{km~s$^{-1}$}
\newcommand{\ms}{m~s$^{-1}$}
\newcommand{\etal}{{et al.}}
\newcommand{\Qm}{$Q_\mu$}

\usepackage[T1]{fontenc}
\usepackage{ae,aecompl}

\title[Methanol isotopologues as a probe for $\mu$-variation]
{\textit{
Methanol isotopologues as a probe for spatial and temporal variations of
the electron-to-proton mass ratio
}}

\author[J. S. Vorotyntseva \etal ] {
J. S. Vorotyntseva$^{1,2}$,
M. G. Kozlov$^{2,3}$,
S. A. Levshakov$^{1,2}$\thanks{E-mail: lev@astro.ioffe.ru}
\vspace*{8pt}
\\
$^{1}$Ioffe Physical-Technical Institute, 194021 St.~Petersburg, Russia\\
$^{2}$Department of Physics,
Electrotechnical University ``LETI'', 197376 St.~Petersburg, Russia\\
$^{3}$Petersburg Nuclear Physics Institute of NRC "Kurchatov Institute", 
Gatchina, Leningrad District, 188300, Russia\\
}
\date{Accepted  Received ; in original form 2023 September }

\pubyear{2023}

\begin{document}
\label{firstpage}
\pagerange{\pageref{firstpage}--\pageref{lastpage}}
\maketitle

\begin{abstract}
We present results on numerical calculations of the sensitivity coefficients, $Q_\mu$, of microwave
molecular transitions in $^{13}$CH$_3$OH and CH$_3$$^{18}$OH to the hypothetical variation 
in the fundamental physical constant $\mu$~-- the electron-to-proton mass ratio. The invariability
of $\mu$ in time and space is one of the basic assumptions of the Standard Model of particle physics
which can be tested at cosmological scales
by means of astronomical observations in the Galaxy and external galaxies.
Our calculations show that these two methanol isotopologues can be utilized for such tests since their microwave
transitions from the frequency interval 1--100 GHz exhibit a large spread in $Q_\mu$ values which span
a range of $-109 \la Q_\mu \la 78$.   
We show that the thermal emission lines of $^{13}$CH$_3$OH observed in the star-forming region NGC~6334I
constrain the variability of $\mu$ at a level of 
$3\times10^{-8}$ $(1\sigma)$,
which is in line with the most stringent upper limits
obtained previously from observations of methanol (CH$_3$OH) and other 
molecules in the Galaxy.
\end{abstract}

\begin{keywords}
methods: numerical --
techniques: spectroscopic --
ISM: molecules -- 
ISM: individual objects: NGC~6334I --
elementary particles  
\end{keywords}


\section{Introduction}
\label{Sec1}

Physical theories extending the Standard Model (SM) of elementary particle physics allow for possible
space-time variations of fundamental physical constants. 
Modern experimental capabilities make it possible to test such theories with high
accuracy both in laboratory experiments with atomic clocks and in studies of astronomical objects.
The main research in this area has focused on dimensionless
constants such as the fine
structure constant $\alpha$=$e^2/\hbar c$ and the electron-to-proton mass ratio $\mu$=$m_e$/$m_p$
since they are independent of the choice of unit system.
Small changes in $\alpha$ and $\mu$ produce offsets in the line positions of atoms
and molecules (for a review, see, e.g., Kozlov \& Levshakov 2013).

Measurements of fractional changes in $\mu$ are based on the facts that ($i$) the
the molecular energy levels are dependent on $\mu$ (Thompson 1975), and that ($ii$)
the molecular electron-vibro-rotational transitions have specific dependencies on $\mu$  
(Varshalovich \& Levshakov 1993). The response of a transition to a variation
of $\mu$ is characterized by its dimensionless sensitivity coefficient \Qm, which is
defined as 
\begin{equation}
Q_\mu = \frac{df/f}{d\mu/\mu}\ ,
\label{Eq1}
\end{equation}
where the fractional changes in the electron-to-proton
mass ratio is given by
\begin{equation}
{\Delta\mu}/{\mu}= (\mu_{\rm obs}-\mu_{\rm lab})/\mu_{\rm lab}\ .
\label{Eq2}
\end{equation}
Here $\mu_{\rm obs}$ and $\mu_{\rm lab}$ are the extraterrestrial and terrestrial values of $\mu$, respectively.

In molecules, enhanced sensitivity coefficients are found in tunneling transitions since
probability of tunneling depends exponentially on the mass of tunneling particles.
The effect was firstly considered for the inversion transition of NH$_3$ by Flambaum \& Kozlov (2007)
and later on for hindered rotations in the non-rigid tops (Jansen \etal\ 2011; Levshakov \etal\ 2011).
The tunneling occurs generally in every internal rotor molecule, especially in methanol (CH$_3$OH)
where the methyl group CH$_3$ can make torsional vibrations
with respect to the hydroxyl group OH.
The hydrogen atom in the hydroxyl group can be placed in three
possible positions with equal energies, and in order to move from one configuration to another, it must pass
through the potential barrier caused by the methyl group. Thus, there is a hindered internal rotation
(rotation of the hydrogen atom with respect to the methyl group).
This type of molecules are also characterized by a strong interaction
between internal (hindered) and overall rotations.

The sensitivity coefficients for such transitions in various molecules were previously calculated
in a series of papers.
For example, the sensitivity coefficient of
the $(J,K) = (1,1)$ inversion transition of ammonia (NH$_3$) is equal to
$K_\mu = -4.46$\footnote{It should be noted, that $Q_\mu=-K_\mu$,
since $K_\mu$ is used when $\mu$ is defined as the $proton$-$to$-$electron$ mass ratio, while in our
concept, $\mu$ is the $electron$-$to$-$proton$ mass ratio.}
(Flambaum \& Kozlov 2007),
in the methyl mercaptan molecule CH$_3$SH the sensitivity coefficients 
range from $K_\mu = -14.8$ to 12.2 (Jansen \etal\ 2013), in the methylamine molecule CH$_3$NH$_2$~--
from $ K_\mu = -19$ to  $K_\mu = 24$ (Ilyushin \etal\ 2012). 
The calculations of the sensitivity coefficients for methanol CH$_3$OH were performed by two independent methods
(Jansen \etal\ 2011, Levshakov \etal\ 2011) resulting in the self-consistent
values of \Qm\ ranging from $-17$ to 43.

Astrophysical limits on \dmm\ have been obtained by different methods
in observations of Galactic and
extragalactic objects.
The extragalactic constraint on the variability of $\mu$ at a level of $8\times10^{-6}$ ($1\sigma$)
at the highest redshift $z = 4.22$ (look-back time 12.4 Gyr) towards the quasar J1443+2724 is
deduced from the analysis of the Lyman and Werner absorption lines of molecular hydrogen H$_2$
(Bagdonaite \etal\ 2015).
In the Milky Way, the most stringent upper limits on $\mu$-variations  
based on ammonia and methanol 
observations are, correspondingly, $\Delta\mu/\mu < 2\times10^{-8}$ 
(Levshakov \etal\ 2013), and $\Delta\mu/\mu < 6\times10^{-8}$ (Dapr\`a \etal\ 2017).
Both limits are given at a $3\sigma$ confidence level.
However, it should be noted that these constraints were obtained at the marginal
accuracy of modern spectral observations and, therefore, may be affected by systematics
of unknown values and caused by inaccessible factors.
In order to estimate such kind of systematics, measurements involving different facilities,
different objects, and different probes~-- in our case molecular transitions~-- are required.

Here we calculate the sensitivity coefficients for methanol isotopologues
$^{13}$CH$_3$OH and CH$_3$$^{18}$OH in order to check whether they are suitable to test
$\mu$-variations at the above mentioned levels. 
Section 2 describes the effective Hamiltonian and
its parameters, as well as the determination of the sensitivity coefficients.
Section 3 presents the results of these calculations. 
In Section 4, we describe existing detections of isotopic methanol transitions
and use observations of the $^{13}$CH$_3$OH thermal 
emission lines from the young stellar object NGC~6334I
to constrain \dmm. 
The results obtained are summarized in Section 5.

\section{Calculating procedure}
\label{Sec2}

\subsection{Quantum-mechanical model}
\label{SSec21}

The procedure for calculating sensitivity coefficients is described in details in Levshakov \etal\ (2011). Here are given only the main aspects.
The procedure is based on an approach by Rabli \& Flower (2010) which implies a simple
and convenient form of the effective Hamiltonian with six spectroscopic constants having
clear physical meaning. The dependence of these spectroscopic constants
on $\mu$ is easily understood within the Born-Oppenheimer approximation.
This Hamiltonian is physically transparent and quite accurate for calculations of sensitivity coefficients.

In Rabli \& Flower (2010), it is applied a direct method of solving the Schr$\rm\ddot o$dinger equation
which involves diagonalizing the Hamiltonian,
\begin{equation}
\hat{H} = -F\frac{d^2}{d\omega^2}+V(\omega),
\label{Eq4}
\end{equation}
expressed on a basis of torsion-rotation states
\begin{equation}
|JKm\rangle=\left(\frac{2J+1}{16\pi^3}\right)^{1/2}D_{Km_J}^Je^{-i\rho K\omega}e^{-im\omega},
\label{Eq5}
\end{equation}
where $0 \le \omega \le 2\pi$ is the torsion angle of the internal rotation of the CH$_3$ group
relative to the OH radical, $V(\omega)$ is the torsional potential, $F$ is the torsional constant (see below),
$D_{Km_J}^J$ is a rotation matrix element which is a function of the Euler angles specifying the orientation of
the molecule relative to a space-fixed laboratory coordinate system, $J$ is the rotational angular momentum, $m_J$ is the projection of the rotational angular momentum on the space-fixed $z$-axis,
$K$ is its projection on the symmetry axis of the molecule, $\rho = 0.8102$ for $^{13}$CH$_3$OH (Xu \& Lovas 1997), and
$\rho = 0.809$ for CH$_3$$^{18}$OH  (Fisher \etal\ 2007) is the fraction contributed by the CH$_3$ group to
the rotational angular momentum of the molecule about its symmetry axis.

For a given value of $J$, the non-vanishing matrix elements of the methanol Hamiltonian, expressed
in the mentioned above basis, are the following 
(in this model the conversion for the rotational parameters $A>B>C$ were used)
\begin{equation}
\begin{array}{l}
\langle JKm'|\hat{H}|JKm\rangle = \delta_{m',m} \times \\[5pt]
\times \left(\frac{B+C}{2}[J(J+1)-K^2]+AK^2+F(m-\rho K)^2+\frac{V_3}{2}\right) - \\[5pt]
- \delta_{|m'-m|,3}\frac{V_3}{4}\ ,
\end{array}
\label{Eq6}
\end{equation}
\begin{equation}
\begin{array}{l}
\langle J K \pm 2, m'|\hat{H}|JKm\rangle = \delta_{m',m}[J(J+1)-K(K \pm 1)]^{1/2}\times \\[5pt]
\times \frac{B-C}{4}[J(J+1)-(K \pm 1)(K \pm 2)]^{1/2}\ ,\\
\end{array}
\label{Eq7}
\end{equation}
\begin{equation}
\begin{array}{l}
\langle JK \pm 1, m'|\hat{H}|JKm \rangle = \delta_{m',m} [J(J+1)-K(K \pm 1)]^{1/2} \times \\[5pt]
\times \frac{D}{2}(2K \pm 1)\ ,
\end{array}
\label{Eq8}
\end{equation}
where $\delta_{i,k}$ is Kronecker's symbol.
Diagonalization of the Hamiltonian matrix, formed by these matrix elements,
yields the eigen-energies and eigenfunctions of $A$- and $E$-methanol simultaneously, for a given $J$.
The torsional angular momentum quantum number, $m$, satisfies the relation $m = 3s + \sigma$, where
$s$ is any integer and $\sigma = 0$ for $A$-methanol, $\sigma= \pm 1$ for $E$-methanol.

Six parameters of the effective Hamiltonian are following: three rotational parameters $A$, $B$, $C$, 
one parameter $D$, describing interaction of internal rotation with overall rotation,
the kinetic coefficient $F$, and the depth of the three-fold symmetric torsion potential $V_3$,
\begin{equation}
V(\omega)=\frac{V_3}{2}(1 - \cos3\omega).
\label{Eq9}
\end{equation}
The values of $V_3$ and $F$ are taken from Xu \& Lovas (1997) for $^{13}$CH$_3$OH 
and Fisher \etal\ (2007) for CH$_3$$^{18}$OH,
parameters $A$, $B$, $C$, $D$ were calculated from the moments of inertia of $^{13}$CH$_3$OH
and CH$_3$$^{18}$OH (Lees \etal\ 1973) according to the formulas:
\begin{equation}
A = \frac{1}{2}\hbar^2 \left(\frac{I_a+I_b}{I_aI_b-I^2_{ab}}-\frac{I_b}{I^2_b+I^2_{ab}}\right),
\label{Eq10}
\end{equation}

\begin{equation}
B = \frac{1}{2}\hbar^2 \frac{I_b}{I^2_b+I^2_{ab}},
\label{Eq11}
\end{equation}

\begin{equation}
C = \frac{1}{2}\hbar^2 \frac{1}{I_c},
\label{Eq12}
\end{equation}

\begin{equation}
D = \frac{1}{2}\hbar^2 \frac{I_{ab}}{I^2_b+I^2_{ab}}.
\label{Eq13}
\end{equation}

\smallskip\noindent
In these equations, $\hbar=h/2\pi$, $I_a$, $I_b$, $I_c$ are the moments of inertia, and $I_{ab}$
is the product of inertia about the $a$- and $b$-axes in the $a,b,c$-axis system whose $a$-axis is parallel
to the internal rotation axis (assumed to be that of the methyl top), the $c$-axis perpendicular to the COH plane
(for details, see Lees \& Baker 1968). 
The numerical values of the moments of inertia are given in Table~\ref{T1}, while 
Table~\ref{T2} lists the calculated spectroscopic parameters for $^{13}$CH$_3$OH, CH$_3$$^{18}$OH,
and $^{12}$CH$_3$$^{16}$OH 
along with the kinetic coefficients $F$ and potential barriers $V_3$.

Errors of $I_a$, $I_b$, $I_c, I_{ab}$ and $A, B, C, D$
are not given in Tables~\ref{T1} and \ref{T2}, since, in general,
errors in \Qm\ caused by the model itself are by orders of magnitude
larger than uncertainties in the moments of inertia and in 
the spectroscopic rotational parameters. Therefore, in what follows we
consider these values as constants.

\subsection{Determination of the sensitivity coefficients}
\label{SSec22}

Microwave transitions, in our case, torsion-rotation transitions, can be expressed by the quantum numbers
of the upper and lower levels~-- total angular momentum $J$ and its projection $K$ on the axis of the molecule.
Hereafter, it will be denoted as $J_i$ and $K_i$~-- quantum numbers for the upper level,
and $J_k$ and $K_k$~-- quantum numbers for the lower level.

The sensitivity coefficient $Q_{\mu,ik}$ for the transition $J_iK_i \to J_kK_k$ is given by
\begin{equation}
Q_{\mu,ik}={q_{ik}}/{f},
\label{Eq14}
\end{equation}
where $q_{ik} = q_i - q_k$, $q_i$ and $q_k$ are the $q$-factors, individual for each level,
which shows a response of the level to a small change of $\mu$, and $f$ is the laboratory transition
frequency taken from 
Anderson \etal\ (1987, 1990), Kuriyama \etal\ (1986),
Hughes \etal\ (1951) for $^{13}$CH$_3$OH, and from Ikeda \etal\ (1998) for CH$_3$$^{18}$OH.

To determine the sensitivity coefficients \Qm, 
we first find the dependence of the eigenvalues $E_i$ on $\Delta\mu/\mu$:
\begin{equation}
\Delta E_i = q_i{\Delta\mu}/{\mu},
\label{Eq15}
\end{equation}
where the coefficient $q_i$ reveals the response 
of the level $E_i$ to a small change in $\mu$, i.e., 
$|\Delta\mu/\mu| \ll 1$. This is done by diagonalizing the effective 
Hamiltonian for the three sets of parameters that correspond to 
$\mu = \mu_0$ and $\mu = \mu_0(1 \pm \varepsilon)$, where $\varepsilon$ 
is equal to 0.001 or 0.0001 (for details, see Levshakov \etal\ 2011).

\section{Results}
\label{Sec3}

Molecular transitions with the total angular momentum $J$ from 0 to 13 were selected for calculations,
and the quantum numbers $K$ for the upper and lower levels were chosen for $|\Delta K| = 1$,
since transitions without changing $K$ are purely rotational 
and have approximately the same sensitivity to $\mu$-variations, $Q_{rot} \approx 1$.
The ranges of $J$ values chosen are based on the fact that such transitions are described by
our model quite well. For $J > 13$, the model turns out to be insufficiently accurate.
Transitions of a wide frequency interval from 1 to 900 GHz are considered.
Among them low-frequency transitions are found to be the most interesting, since they have
the highest sensitivities to changes in $\mu$.

The calculated sensitivity coefficients for $^{13}$CH$_3$OH are listed in Table~\ref{T3}
where the quantum numbers $J$ and $K$ of the upper and lower levels, the rest frequency
$f$, and the sensitivity coefficient \Qm\ are indicated.
Table~\ref{T4} contains the similar information as Table~\ref{T3} but for CH$_3$$^{18}$OH.
Both tables show only transitions with large sensitivity coefficients, $|Q_\mu| \gg 1$.
These transitions can be observed under real conditions in the interstellar medium.

Errors of the sensitivity coefficients were calculated as follows. 
Parameters of the spin-rotational Hamiltonian depend on the mass ratio $\mu$.
One can see from Eqs.~(\ref{Eq10}-\ref{Eq13})
that $A$, $B$, $C$, and $D$ are inversely proportional to the molecular moments of inertia.
To a first approximation, the equilibrium internuclear distances do not depend on $\mu$ and, 
therefore, these parameters scale linearly with $\mu$.
Within this approximation the kinetic coefficient $F$ in Eq.~(\ref{Eq4}) is also proportional 
to $\mu$, 
while the potential barrier $V_3$ in Eq.~(\ref{Eq9}) is independent of $\mu$.
However, the vibrational wave functions of the molecule depend on $\mu$ and in the next approximation 
the internuclear distances also weakly depend on $\mu$.
This, as well as centrifugal and other corrections, affects scaling of all parameters with $\mu$.
We estimated in Levshakov \etal\ (2011) that the Hamiltonian parameters are proportional 
to $\mu^{n+\varepsilon}$, where $n=0$ for $V_3$ and $n=1$ for all other parameters, while $|\varepsilon|<0.02$.
This uncertainty is the dominant source of errors when we calculate sensitivity coefficients $Q_\mu$. 
The resulting errors are determined by summing quadratically the errors caused by changes in the 
scalings of these six parameters within the uncertainty interval $|\varepsilon|<0.02$.
The errors determined in this way are majorizing estimates of the uncertainties in the values
of the sensitivity coefficients \Qm.

We note that our procedure yields \Qm\ in concordance with values
for two lines of CH$_3$$^{18}$OH at 2.604 GHz and 11.629 GHz,
and one line of $^{13}$CH$_3$OH at 1.989 GHz found by Jansen \etal\ (2011), who used
a different numerical approach. 

As Table~\ref{T3} shows, there are significantly different sensitivity coefficients
in $^{13}$CH$_3$OH with values for $E$-methanol span an interval from $Q_\mu = -32$ for the
$10_{-4} - 11_{-3}E$ line at 9.999 GHz to $Q_\mu = 78$ for the 
$8_{-2} - 9_{-1}E$ line at 1.989 GHz.
For $A$-methanol, \Qm\ ranges from $Q_\mu = -31$ for the 
$10_{-1} - 9_2A^-$ line at 9.153 GHz to $Q_\mu = 21$ for 
the $5_1 - 6_0A^+$ line at 14.300 GHz.

For CH$_3$$^{18}$OH, as Table~\ref{T4} shows, the most notable are the transitions
$10_1 - 9_{2}A^-$ (2.604 GHz) with $Q_\mu = -109$, and
$5_1 - 6_{0}A^+$ (15.134 GHz) with $Q_\mu = 19$, and a series of transitions
$J_2 - J_1$ (with $J=2-13$) of $E$-methanol ($34-36$ GHz) with \Qm\ between $-12$ and $-10$.

Thus, methanol isotopologues $^{13}$CH$_3$OH and CH$_3$$^{18}$OH can provide of
a distinct interest in the issues of $\mu$-variations:
\begin{itemize}
\item there are significantly different sensitivity coefficients of both signs between different transitions
with $\Delta K = \pm1$;
\item all transitions with $\Delta K = 0$ have $Q_\mu \approx 1$, and thus can serve as anchors together
with highly sensitive lines;
\item these transitions are in a fairly narrow frequency range which makes them convenient for 
simultaneous observations;
\item methanol isotopologues, being observed at various galactocentric distances,
can be used for probing 
the hypothetical coupling of the dark matter with the baryonic matter depending on
local environmental conditions.
\end{itemize}

\section{Observational constraints on $\mu$-variations }
\label{Sec4}

Methanol isotopologues have previously been observed in Galactic
and extragalactic sources at different frequencies.
In most cases, high-frequency transitions were observed for both isotopologues (e.g., Nummelin \etal\ 1998).
The most common objects with the methanol isotopologues emission are located in the
molecular cloud complex in the Orion (OMC-1) and in the molecular cloud near the center of the Galaxy (Sgr~B2).

In addition to these two locations, the {\it emission lines} of $^{13}$CH$_3$OH were detected
in the young stellar object (YSO) IRAS16293-2422 (Parise \etal\ 2004; Kuan \etal\ 2004),
in the star-forming region G34.3+0.15 (Macdonald \etal\ 1996), and in the HII region IRAS18089-1732
(Beuither \etal\ 2004), whereas the {\it absorption lines} of $^{13}$CH$_3$OH
were previously observed towards the continuum sources Sgr~A, Sgr~B2, and W33
(Kuiper \etal\ 1989) and in one of the two lines of sight towards
the quasar PKS~1830-211 at redshift $z = 0.89$ 
(Muller \etal\ 2021; Kanekar \etal\ 2015; Bagdonaite \etal\ 2013).

In this extragalactic absorber,
the detected high-frequency transitions of $^{13}$CH$_3$OH were used
to investigate the cosmological invariance of $\mu$ at a look-back time of half the present age
of the Universe.
For this purpose the authors calculated sensitivity coefficients for
the  transitions $1_{-1} - 1_{0}A$ at 303.692 GHz, 
$2_{-1} - 2_{0}A$ at 304.494 GHz, and 
$3_{-1} - 3_{0}A$ at 305.699 GHz
($Q_\mu = 1.904, 1.902$, and 1.898, respectively) and estimated the upper limit on
$|\Delta \mu/\mu| < 1.2\times10^{-7}$ ($1\sigma$).
However, as noted above, high-frequency transitions do not have large values of \Qm\
making high-precision measurements of \dmm\ difficult.

But a remarkable fact in methanol observations is that
its isotopologues have also been observed in the Galaxy
at low-frequencies with high \Qm\ values of different signs which is of great interest for
probing spatial and temporal variations of $\mu$.

For instance, a series of {\it thermal} emission lines of $^{13}$CH$_3$OH (23-44 GHz)
was recently observed in the YSO NGC~6334I by Wu \etal\ (2023).
We selected from their Table~3 the most accurate line positions with known
sensitivity coefficients and listed them in our Table~\ref{T5}.
It is seen that $\Delta Q_\mu$ in this case can be as large as 20.6.

The fractional changes in $\mu$ can be estimated from a pair of molecular transitions ($i$, $j$)
with different values of \Qm\ (Levshakov \etal\ 2022):
\begin{equation}
\frac{\Delta\mu}{\mu}= \frac{V_j-V_i}{c(Q_{\mu,i}-Q_{\mu,j})},
\label{Eq3a}
\end{equation}
where $V_j$ and $V_i$ are the local standard of rest radial velocities, $V_{\rm LSR}$,
of molecular transitions with corresponding sensitivity coefficients $Q_{\mu,j}$ and 
$Q_{\mu,i}$, and $c$ is the speed of light.

The sample mean \dmm\ and its error based on the total list of $n = 8$ lines from Table~\ref{T5}
is expected to be more precise than that based on a single pair of lines with the largest difference
between \Qm\ values, e.g., between the $5th$ and $1st$ lines.
However, the improvement is not as high as $1/\sqrt{n}$ because
the individual values of \dmm\ are correlated.

Namely, from a set of $n$ radial velocities $\{ V_1, V_2, \ldots, V_n\}$ we can form $(n-1)$
velocity differences taking the first of them with positive \Qm\ as a reference velocity:
$\{ V_2-V_1, V_3-V_1, \ldots, V_n-V_1\}$.
From this dataset we can form, in turn, $(n-1)$ values of \dmm, using Eq.(\ref{Eq3a}):
$\{$(\dmm)$_1$, (\dmm)$_2$,  \ldots, (\dmm)$_{n-1}$$\}$.
Then the correlation coefficient $\kappa_{i,j}$ between two \dmm\ values ($i \neq j$) is given by
(Levshakov \etal\ 2010):
\begin{equation}
\kappa_{i,j} = \left[ (1+s^2_i)(1+s^2_j) \right]^{-1/2}\, ,
\label{Lq15}
\end{equation}
where $s^2_i = \sigma^2_{(\Delta\mu/\mu)_i}/\sigma^2$, $s^2_j = \sigma^2_{(\Delta\mu/\mu)_j}/\sigma^2$,
and $\sigma^2$ is the variance of the most precise estimate of \dmm.
Taking into account that the errors of (\dmm)$_i$, $\sigma_{(\Delta\mu/\mu)_i}$,
are almost equal, we have $\kappa_{i,j} = \kappa \approx 1/2$.

The covariance matrix $Cov$[(\dmm)$_i$,(\dmm)$_j$]
contains $(n-1)$ diagonal terms $\sigma^2$, and $(n-1)(n-2)$
non-diagonal terms $\kappa\sigma^2$.
Then the error of the mean \dmm\ is given by
\begin{equation}
\sigma_{\Delta\mu/\mu} = \left[ \sum^{n-1}_{i=1}\sum^{n-1}_{j=1} w_i w_j
Cov[(\Delta\mu/\mu)_i,(\Delta\mu/\mu)_j] \right]^{1/2}\, .
\label{Lq16}
\end{equation}
In case of equal accuracy, the weight $w_i = 1/(n-1)$ for each $i$, and thus
\begin{equation}
\sigma_{\Delta\mu/\mu} = \frac{\sigma}{(n-1)}\sqrt{(n-1) + (n-1)(n-2)\kappa} \approx \sigma\sqrt{\kappa}\ .
\label{Lq17}
\end{equation}
This implies that the gain factor $\sqrt{\kappa} \approx 0.7$ for this dataset.

With $\sigma = 3.7\times10^{-8}$, based on the $V_1$ and $V_5$ radial velocities,
one obtains for the sample mean \dmm\ the value
$\langle \Delta\mu/\mu \rangle = (3 \pm 3)\times10^{-8}$, 
which is consistent with no variation of $\mu$ at a level of
$3\times10^{-8}$ ($1\sigma$).

Another estimate of $\langle \Delta\mu/\mu \rangle$ and its error can be obtained by
averaging the radial velocities of lines 2--8, which have approximately equal 
sensitivity coefficients, and comparing the result with line 1. In this way, we have
the weighted mean $\langle V \rangle = -7.36\pm0.04$ \kms, 
$\langle Q_\mu \rangle = -14.7\pm0.2$, and 
$\langle \Delta\mu/\mu \rangle = (4\pm3)\times10^{-8}$ ($1\sigma$). 

The obtained constraint is in line with the most stringent upper limit on
$|\Delta \mu/\mu| < 2\times10^{-8}$ ($1\sigma$) found from 
observations of Class~I methanol (CH$_3$OH) masers distributed in the Milky Way disk 
over a large range of the galactocentric distances (Levshakov \etal\ 2022),
and from measurements of methanol thermal emission lines towards 
the dense dark cloud core L1498 (Dapr\`a \etal\ 2017). 

We note in passing that isotopic methanol ($^{13}$CH$_3$OH) {\it maser} emission was recently detected
in the star-forming region G358.93$-$0.03 by Chen \etal\ (2020).
These are two low-frequency transitions: $2_0 - 3_{-1}E$
at 14.782 GHz ($Q_\mu = 27$) and  $5_1 - 6_0A^+$ at 14.300 GHz ($Q_\mu = 21$).

CH$_3$$^{18}$OH was not detected in maser emission, but its low-frequency transitions with high
sensitivity coefficients were also observed in interstellar space.
The {\it emission} line $10_{2} - 10_{1}E$ at 34.831 GHz ($Q_\mu = -11$)
was detected towards NGC~6334I (Wu \etal 2023),
and the {\it absorption} line $2_0 - 3_{-1}E$ at 11.629 GHz ($Q_\mu = 33$)~-- at two positions
in the direction of Sgr~B2 (Gardner \etal\ 1989).

These observations and detections of isotopic methanol transitions
in the low-frequency range from 1 to 40 GHz, where the calculated sensitivity coefficients
demonstrate large values of different signs, show great potential for investigations of
fundamental physical principles on Galactic and extragalactic scales.

\section{Summary}
\label{Sect5}

The numerical calculations discussed in this paper were designed to study response of the microwave molecular 
line positions in the methanol isotopologues $^{13}$CH$_3$OH and CH$_3$$^{18}$OH
on small changes in the electron-to-proton mass ratio, $\mu$. Varying $\mu$ would cause shifts in the line 
positions different for individual lines. This opens up the opportunity to test invariability of $\mu$ 
through comparison of astronomical spectra with laboratory determinations. 

That the electron-to-proton mass ratio may not be a constant would imply that the relative strength of the 
electromagnetic force compared to the strong nuclear force is space-time dependent. No evidences for either spatial 
or temporal changes in $\mu$ have been found yet, however, at a level of a few times $10^{-8}$. 

The improvement in this constraint is currently limited to uncertainties of $\sim 10$ kHz
in the rest frame frequencies of the methanol isotopologues, corresponding to uncertainties 
of $\sim 100$ \ms\ in the velocity scale for the low-frequency microwave transitions listed 
in Tables~\ref{T3}, and \ref{T4}. 
The recently measured radial velocities of 
$^{13}$CH$_3$OH lines (Wu \etal\ 2023) have similar errors (see Table~\ref{T5}).   
Future laboratory and astronomical investigations of methanol isotopologues with higher quality spectra 
can improve by more than 10 times the most stringent limit on $\mu$-variation in the Galaxy
enabling these tests to be performed at a level of $10^{-9}$. 

\smallskip\noindent
Our main results are as follows.
\begin{enumerate}
\item The spectroscopic rotational parameters $A, B, C,$ and $D$ for methanol isotopologues
$^{13}$CH$_3$OH and CH$_3$$^{18}$OH were calculated and presented in Table~\ref{T2}.
\item The previously developed procedure for calculating sensitivity coefficients \Qm\
for methanol lines (Levshakov \etal\ 2011) was used to analyze  
torsion-rotation transitions of the methanol isotopologues in the range 1--100 GHz.
\item The calculated \Qm\ coefficients were shown to have significantly different values
of both signs which span an interval from $-32$ to 78 for $^{13}$CH$_3$OH (Table~\ref{T3}),
and from $-109$  to 19 for CH$_3$$^{18}$OH (Table~\ref{T4}).
\item For the three previously known \Qm~factors in these isotopologues (Jansen \etal\ 2011),
good agreement with our results was obtained.
\item The constraint on the $\mu$-variability at the $3\times10^{-8}$ ($1\sigma$) level was obtained
from the $^{13}$CH$_3$OH thermal emission lines observed recently in the young stellar object
NGC~6334I by Wu \etal\ (2023).
\end{enumerate}

In conclusion, we note that the methanol isotopologues lines were observed in the Milky Way
both near the Galactic nucleus and in the disk at different galactocentric distances, which
opens new possibilities for testing $\mu$ as a function of the gravitational potential on
galactic scales.

\section*{Acknowledgments}

The authors thank Irina Agafonova for useful comments and suggestions. 
We also thank an anonymous referee for diligent reading of our manuscript
and useful remarks.
Supported in part by the Russian Science Foundation
under grant No.~23-22-00124.

\section*{Data Availability}

The data underlying this article will be shared on reasonable request to the
corresponding author.

\begin{table}
\centering
\caption{Moments of inertia for the main methanol molecule and its isotopologues
(in units amu~\AA$^2$).}
\label{T1}
\begin{tabular}{l r@.l r@.l r@.l}
\hline\\[-5pt]
 &\multicolumn{2}{c}{$^{12}$CH$_3$$^{16}$OH$^a$}& \multicolumn{2}{c}{$^{13}$CH$_3$OH$^b$} 
& \multicolumn{2}{c}{CH$_3$$^{18}$OH$^b$}\\
\hline\\[-7pt]
$I_{a}$ & 3&96277 &3&9618     &3&9697\\
$I_{b}$ & 20&4834 &20&993    &21&381\\
$I_{ab}$ & $-0$&065 &$-0$&076  &$-0$&151\\
$I_{c}$ & 21&2679 &21&777     &22&173\\
\hline\\[-8pt]
\multicolumn{2}{l}{\footnotesize  {\it References:}} & \multicolumn{5}{l}{\footnotesize  $^a$Lees \& Baker (1968); }\\
\multicolumn{2}{l}{} &  \multicolumn{5}{l}{\footnotesize  $^b$Lees \etal\ (1973).}
\end{tabular}
\end{table}

\begin{table}
\centering
\caption{Parameters of the effective Hamiltonian (in units cm$^{-1}$)
for $^{13}$CH$_3$OH, CH$_3$$^{18}$OH, and 
$^{12}$CH$_3$$^{16}$OH, for comparison.}
\label{T2}
\begin{tabular}{ c r@.l r@.l r@.l }
\hline\\[-5pt]
 & \multicolumn{2}{c}{$^{13}$CH$_3$OH} & \multicolumn{2}{c}{CH$_3$$^{18}$OH}
& \multicolumn{2}{c}{$^{12}$CH$_3$$^{16}$OH}   \\
\hline\\[-7pt]
$A$ & 4&2555 & 4&2479  & 4&25427\\
$B$ & 0&8030  & 0&78840   & 0&82298\\
$C$ & 0&77410  & 0&76028)   & 0&75721\\
$D$ & $-0$&0029   &$-0$&0056    &$-0$&00261)\\
$F$ & 27&641920$^a$ & 27&4284105$^b$ &27&646819$^a$\\
$V_3$ & 373&77677$^a$ & 374&06655$^b$ & 373&594$^a$\\
\hline\\[-8pt]
\multicolumn{7}{l}{\footnotesize  {\it References:} $^a$Xu \& Lovas (1997); $^b$Fisher \etal\ (2007)}
\end{tabular}
\end{table}

\begin{table}
\caption{Calculated sensitivity coefficients $Q_\mu$
for the torsion-rotation transitions ($\Delta K = \pm 1$) in $^{13}$CH$_3$OH.
Given in parentheses are errors in the last digits.}
\label{T3}
\begin{tabular}{l r@.l  r@.l }
\hline\\[-5pt]
\multicolumn{1}{l}{Transition} & \multicolumn{2}{c}{Frequency, $f^a$} & 
\multicolumn{2}{c}{\hspace{1.1cm} $Q_\mu$} \\
\multicolumn{1}{c}{$J_{k_{K_k}} - J_{i_{K_i}}$ } & \multicolumn{2}{c}{(MHz)} \\
\hline\\[-7pt]
{$8_{-2}-9_{-1}E$}& 1989&502 & \multicolumn{2}{r}{78(6)$^b$}\\
{$10_{1}-9_{2}A^-$}& 9153&500 & \multicolumn{2}{r}{$-$31(2)}\\
{$10_{-4}-11_{-3}E$}& 9999&400 & \multicolumn{2}{r}{$-$32(2)}\\
{$5_{1}-6_{0}A^+$}& 14300&350 & 20&6(9)\\
{$2_{0}-3_{-1}E$}& 14782&270 & 26&8(1.0)\\
{$4_{3}-5_{2}A^+$}& 19123&400 & $-$21&3(1.1)\\
{$4_{3}-5_{2}A^-$}& 19195&450 & $-$21&3(1.1)\\
{$2_{1}-3_{0}E$}& 23980&250 & 5&4(2)\\
{$3_{2}-3_{1}E$}& 27047&280 & $-$15&2(7)\\
{$4_{2}-4_{1}E$}& 27050&540 & $-$15&2(7)\\
{$2_{2}-2_{1}E$}& 27053&030 & $-$15&2(7)\\
{$5_{2}-5_{1}E$}& 27071&930 & $-$15&2(7)\\
{$6_{2}-6_{1}E$}& 27122&720 & $-$15&1(7)\\
{$7_{2}-7_{1}E$}& 27215&570 & $-$15&0(7)\\
{$8_{2}-8_{1}E$}& 27364&090 & $-$14&9(7)\\
{$9_{2}-9_{1}E$}& 27581&630 & $-$14&7(7)\\
{$10_{2}-10_{1}E$}& 27880&030 & $-$14&5(7)\\
{$6_{2}-5_{3}A^-$}& 27992&990 & 16&3(8)\\
{$6_{2}-5_{3}A^+$}& 28137&250 & 16&2(8)\\
{$11_{2}-11_{1}E$}& 28267&770 & $-$14&2(6)\\
{$12_{2}-12_{1}E$}& 28747&750 & $-$13&8(6)\\
{$13_{2}-13_{1}E$}& 29315&200 & $-$13&4(6)\\
{$7_{0}-6_{1}A^+$}& 35161&580 & $-$7&0(4)\\
{$9_{2}-10_{1}A^+$}& 35171&780 & 9&3(7)\\
{$8_{2}-9_{1}A^-$}& 41904&330 & 7&9(5)\\
{$7_{2}-6_{3}A^-$}& 75155&150 & 6&7(3)\\
{$7_{2}-6_{3}A^+$}& 75415&300 & 6&7(3)\\
\hline\\[-8pt]
\multicolumn{4}{l}{\footnotesize {\it Notes.} 
$^a$The rest frequencies are taken from }\\ 
\multicolumn{4}{l}{\footnotesize Anderson \etal\ (1987, 1990); Kuriyama }\\ 
\multicolumn{4}{l}{\footnotesize \etal\ (1986); Hughes \etal\ (1951).}\\
\multicolumn{4}{l}{\footnotesize $^bQ_\mu = 63\pm3$ from Jansen \etal\ (2011).}
\end{tabular}
\end{table}

\begin{table}
\caption{Calculated sensitivity coefficients $Q_\mu$
for the torsion-rotation transitions ($\Delta K = \pm 1$) in CH$_3$$^{18}$OH.
Given in parentheses are errors in the last digits.}
\label{T4}
\begin{tabular}{l r@.l r@.l }
\hline\\[-5pt]
\multicolumn{1}{c}{Transition} & \multicolumn{2}{c}{Frequency, $f^a$} & 
\multicolumn{2}{c}{\hspace{1.1cm} $Q_\mu$} \\
\multicolumn{1}{c}{$J_{k_{K_k}} - J_{i_{K_i}}$ } & \multicolumn{2}{c}{(MHz)} \\
\hline\\[-7pt]
{$10_{1}-9_{2}A^-$}& 2604&912& \multicolumn{2}{r}{$-$109(9)$^b$}\\
{$2_{0}-3_{-1}E$}& 11629&69& 33&4(1.3)$^c$\\
{$5_{1}-6_{0}A^+$}& 15134&717& 19&3(8)\\
{$12_{-3}-11_{-4}E$}& 18880&216& 18&1(1.3)\\
{$2_{1}-3_{0}E$}& 25681&549& 5&1(2)\\
{$10_{-4}-11_{-3}E$}& 27492&895& $-$10&7(9)\\
{$4_{3}-5_{2}A^+$}& 31117&221& $-$12&5(7)\\
{$4_{3}-5_{2}A^-$}& 31186&461& $-$12&5(7)\\
{$7_{0}-6_{1}A^+$}& 33429&897& $-$7&3(4)\\
{$2_{2}-2_{1}E$}& 33919&029& $-$11&8(5)\\
{$3_{2}-3_{1}E$}& 33925&484& $-$11&7(5)\\
{$4_{2}-4_{1}E$}& 33943&605& $-$11&7(5)\\
{$5_{2}-5_{1}E$}& 33981&375& $-$11&7(5)\\
{$6_{2}-6_{1}E$}& 34048&413& $-$11&6(5)\\
{$7_{2}-7_{1}E$}& 34155&632& $-$11&6(5)\\
{$8_{2}-8_{1}E$}& 34314&639& $-$11&5(5)\\
{$9_{2}-9_{1}E$}& 34536&692& $-$11&3(5)\\
{$4_{-1}-3_{0}E$}& 34709&266& $-$9&9(4)\\
{$10_{2}-10_{1}E$}& 34831&635& $-$11&2(5)\\
{$11_{2}-11_{1}E$}& 35206&084& $-$11&0(5)\\
{$12_{2}-12_{1}E$}& 35661&816& $-$10&7(5)\\
{$13_{2}-13_{1}E$}& 36193&552& $-$10&5(5)\\
{$9_{2}-10_{1}A^+$}& 40561&430& 8&1(6)\\
{$8_{2}-9_{1}A^-$}& 47517&950& 7&0(4)\\
{$1_{0}-2_{-1}E$}& 57974&895& 7&5(3)\\
{$7_{2}-6_{3}A^-$}& 61471&770& 7&8(3)\\
{$7_{2}-6_{3}A^+$}& 61721&517& 7&8(3)\\
{$4_{1}-5_{0}A^+$}& 63350&909& 5&4(2)\\
{$13_{-3}-12_{-4}E$}& 65249&808& 5&9(4)\\
{$3_{1}-2_{2}E$}& 105180&779&5&1(2)\\
\hline\\[-8pt]
\multicolumn{4}{l}{\footnotesize {\it Notes.} 
$^a$The rest frequencies are taken from }\\
\multicolumn{4}{l}{\footnotesize Ikeda \etal\ (1998). $^bQ_\mu = -93\pm5$,  }\\
\multicolumn{4}{l}{\footnotesize $^cQ_\mu = 34\pm2$ from Jansen \etal\ (2011). }\\
\end{tabular}
\end{table}

\begin{table}
\caption{$^{13}$CH$_3$OH lines towards NGC~6334I detected
from the Tian Ma 65-m radio telescope observations
by Wu \etal\ (2023).
Shown in columns are ordinal number,
the centroid velocity, $V_{\rm LSR}$, main-beam temperature of the peak, $T_{\rm mb}$,
and the sensitivity coefficient, $Q_\mu$.
Given in parentheses are errors in the last digits.}
\label{T5}
\begin{tabular}{c l c r@.l r@.l }
\hline\\[-5pt]
No. & \multicolumn{1}{c}{Frequency} & $V_{\rm LSR}$ & \multicolumn{2}{c}{$T_{\rm mb}$} & 
\multicolumn{2}{c}{$Q_\mu$}\\
 & (MHz) & (\kms) & \multicolumn{2}{c}{(K)} \\
\hline\\[-7pt]
1 & 23980.222 & $-7.6(2)$ & 0&27(8) & 5&4(2) \\
2 & 27071.93  & $-7.4(2)$ & 0&31(7) & $-15$&2(7) \\
3 & 27122.72  & $-7.4(1)$ & 0&57(10)& $-15$&1(7) \\
4 & 27215.59  & $-7.3(1)$ & 0&56(10)& $-15$&0(7) \\
5 & 27364.077 & $-7.3(1)$ & 0&71(11)& $-14$&9(7) \\
6 & 27581.616 & $-7.3(1)$ & 0&68(8) & $-14$&7(7) \\
7 & 27880.03  & $-7.6(2)$ & 0&45(10)& $-14$&5(7) \\
8 & 28747.709 & $-7.6(2)$ & 0&38(9) & $-13$&8(6) \\
\hline\\[-5pt]
\end{tabular}
\end{table}

\bsp    
\label{lastpage}
\end{document}